**Business Cycles as Collective Risk Fluctuations**

Victor Olkhov

TVEL, Moscow, Russia

victor.olkhov@gmail.com

ORCID iD 0000-0003-0944-5113

## Abstract

We suggest use continuous numerical risk grades [0,1] of $R$ for a single risk or the unit cube in $R^n$ for $n$ risks as the economic domain. We consider risk ratings of economic agents as their coordinates in the economic domain. Economic activity of agents, economic or other factors change agents risk ratings and that cause motion of agents in the economic domain. Aggregations of variables and transactions of individual agents in small volume of economic domain establish the continuous economic media approximation that describes collective variables, transactions and their flows in the economic domain as functions of risk coordinates. Any economic variable $A(t,\mathbf{x})$ defines mean risk $\mathbf{X}_A(t)$ as risk weighted by economic variable $A(t,\mathbf{x})$. Collective flows of economic variables in bounded economic domain fluctuate from secure to risky area and back. These fluctuations of flows cause time oscillations of macroeconomic variables $A(t)$ and their mean risks $\mathbf{X}_A(t)$ in economic domain and are the origin of any business and credit cycles. We derive equations that describe evolution of collective variables, transactions and their flows in the economic domain. As illustration we present simple self-consistent equations of supply-demand cycles that describe fluctuations of supply, demand and their mean risks.

## 1. Introduction

Cycles of economic activity, macro fluctuations of supply and demand, cycles of investment and economic growth, rise and decline of other macroeconomic variables are the most general and most influential macroeconomic processes. Macroeconomic and business cycles are accompanied by phases of economic growth and development but changed with depressions, economic and financial crises. Understanding the laws that govern macro and business cycles may help better measure, manage and support the prosperity periods and reduce losses of the economic crises. Studies of the economic and financial cycles, endogenous and exogenous factors that initiate, increase or decrease amplitudes and frequency of business cycles have a long history and remain among the most important in the economic research.

Mitchel in 1927 mentioned the processes of business cycles as: "In a business cycle, the order of events is crisis, depression, revival, prosperity, and another crisis (or recession)" (Mitchell, 1927, p.79). Since then not too much was added to the treatment of the business cycles. "The incorporation of cyclical phenomena into the system of economic equilibrium with which they are in apparent contradiction, remains the crucial problem of Trade Cycle Theory" (Hayek, 1933, quoted by Lucas, 1995). "Why aggregate variables undergo repeated fluctuations about trend, all of essentially the same character? Prior to Keynes' General Theory, the resolution of this question was regarded as one of the main outstanding challenges to economic research, and attempts to meet this challenge were called business cycle theory" (Lucas, 1995). Questions on properties and modelling of business cycles remain relevant now and will attract researchers for years. "Theories of business cycles should presumably help us to understand the salient characteristics of the observed pervasive and persistent non seasonal fluctuations of the economy" (Zarnowitz, 1992). "One of the most controversial questions in macroeconomics is what explains business-cycle fluctuations?" (Huggett, 2017). Studies of business cycle since Mitchell (1927), Tinbergen (1935) and Schumpeter (1939) were followed by hundreds publications (Lucas, 1980; Kydland and Prescott 1982; 1991; Zarnowitz 1992; Lucas 1995; Diebold and Rudebusch 1999; Bangia, Diebold and Schuermann, 2000; Rebelo 2005; Kiyotaki 2011; Diebold and Yilmaz 2013; Jorda, Schularick and Taylor 2016; Huggett 2017).

It is obvious that one of the most important problems concern the origin, the source, the initial reasons for the business cycle fluctuations and transitions from the development phase to the "crisis, depression" phase. These problems were studied by (Shapiro and Watson, 1988; Shea, 1999; Andrle, Brůha and Solmaz, 2017). The "crisis, depression" phase is always associated with risk growth. Impacts of risk are treated as part of business cycle studies and count numerous publications (Alvarez and Jermann, 1999; Bangia, Diebold and Schuermann, 2000; Tallarini, 2000; Pesaran, Schuermann and Treutler, 2007; Mendoza, Yue, 2012; Christiano, Motto, and Rostagno, 2013). Selected papers on financial risks over 100 years of research (Diebold, 2012) outlines a special chapter on "Financial Risk And The Business Cycle". Risk measurements, concepts, techniques and tools (McNeil, Frey and Embrechts, 2005) and relations to macro modeling (Brunnermeier and Krishnamurthy, 2014) present only top slice of risks and business cycle studies. Economic crises increase risks of most economic agents and agents change their economic activity in a way that decrease their risks. Relations between risks of economic agents and transitions from one cycle phase to another impact dynamics of the business cycles. Studies of such effect and relations of risk assessment and the business cycle development are described by (Tallarini, 2000; Pesaran, Schuermann & Treutler, 2007; Mendoza & Yue, 2012; Diebold, 2012). Economic and financial risks affect economic stability (Huang, Zhou & Zhu, 2009; Nicolò & Lucchetta, 2011) and pricing models (Bollerslev & Zhang, 2003). Numerous publications study the mutual relations between risks and business cycles (Alvarez & Jermann, 1999; Tallarini, 2000; Pesaran, Schuermann & Treutler, 2007; Christiano, Motto & Rostagno, 2013). General equilibrium considers business cycles as transitions from one equilibrium state to another: "The real business cycle theory is a business cycle application of the Arrow-Debreu model, which is the standard general equilibrium theory of market economies." (Kiyotaki 2011). "Business deterministic cycles will be shown to appear in a purely endogenous fashion under laisser faire. Markets will be assumed to clear in the Walrasian sense at every date, and traders will have perfect foresight along the cycles". (Grandmont, 1985). "Real business cycle models view aggregate economic variables as the outcomes of the decisions made by many individual agents acting to maximize their utility subject to production possibilities and resource constraints. More explicitly, real business cycle models ask the question: How do rational maximizing individuals respond over time to changes in the economic environment and what implications do those responses have for the equilibrium outcomes of aggregate variables?" (Plosser, 1989). Most business cycle models follow general economic equilibrium framework (Lucas, 1975; Kydland & Prescott, 1982; 1991; Grandmont, 1985; Mullineux & Dickinson, 1992; Farmer & Woodford, 1997; Kiyotaki, 2011; Bilbiie, Ghironi & Melitz, 2012; Mendoza & Yue, 2012; Growiec, McAdam & Mućk, 2015; Engle, 2017). Meanwhile remark by Hayek that "cyclical phenomena are in apparent contradiction with economic equilibrium" (Hayek, 1933, quoted by Lucas, 1995) and variability of business cycle properties arise questions that description of the business cycles may need new approaches beyond the general equilibrium concept.

We outline three interrelated economic issues: business cycles, accounting and risk assessment. Business cycles describe fluctuations of macroeconomic and financial variables as economic growth, supply and demand, investment and profits and etc. All macro variables and macroeconomic data are determined by and are depend on accounting data of separate economic agents – banks, corporations, companies and etc. Principles and accuracy of corporate accounting define the accuracy of macroeconomic and financial variables and impact the accuracy on national accounts. On the other hand corporate accounting is heavily relies on risk assessment. Correct corporate accounting must take into consideration impact of intra-corporate risks and action of macroeconomic, market, financial and other economic and financial risks those establish corporate risk environment. Macroeconomic risks change during transitions from one phase of business cycle to another. Thus evolution of business cycles affects risks of economic agents and hence causes variations of corporate and bank accounting. Indeed, assessments of corporate assets depend on agents risk assessments. Hence rise or decline of risks induces opposite asset valuation – decline of asset value with risk growth and rise of value with mitigating risks. The evolution of the business cycles depends on strong feedback of accounting data, asset valuation, agents risks and many other factors.

In (Olkhov, 2016-2019c) we developed successive approximations that describe evolution of macroeconomic variables, transactions and expectations. We consider pure theoretical models. As the central property of the economic evolution we consider the collective risk ratings. We show that macroeconomic variables as supply and demand, investment and credits, profits and consumption and etc., are associated with corresponding collective mean risks. Macroeconomic evolution is accompanied by motion of numerous mean risks related to different macro variables. We show that cyclical macroeconomic development that is usually considered as business cycles is escorted with fluctuations of corresponding mean risks. The initial reasons, political or economic factors, other shocks that that can govern business cycle may be different. But any business cycle is characterized by fluctuations of mean risks. As illustration of this statement we present example that describes cycles of supply and demand that are accompanied with fluctuations of mean supply and mean demand risks. Econometrics, measurements, modeling and forecasting of mean risk fluctuations can be a universal tool for business cycle management. Complexity of relations between business cycle and mean risk fluctuations corresponds the complex nature of macroeconomic processes.

In Sec. 2 we explain the meaning of the economic domain and introduce main assumptions that use agents risk ratings as their coordinates. In Sec 3 we discuss main notions that describe the continuous economic media approximation in the economic domain. In Sec 4 we explain how simple assumptions on economic interactions may model motion of mean risks in economic domain and why this motion cause business cycle fluctuations. We collect almost all formulas into Appendices A, B, C. In Appendices we give formal definitions of economic notions in the economic domain and derive the equations that govern the business cycles. We use roman letters *A, b, c* to define scalars and bold ***X, y, v*** to define vectors. Reference (3.5) means equation 5 in the Section 3.

## 2. Main assumptions

We regard economics as a set of economic agents those perform various market trades. Agent-based economic modeling studied by many researchers and we refer Poggio et.al. (1999), Tesfatsion and Judd (2006), LeBaron and Tesfatsion (2008) and Silverman (2018) to outline that agent-based models (ABM) are well known and are under active research for last decades. We regard all economic entities like multinational corporations, hedge funds, largest banks, small companies and even households as economic agents. Agents have many economic and financial properties those establish the macroeconomic and financial variables: investment and credits, asset and debts, profits and taxes, production function and etc. Some economic variables are additive. For example, sum of credits, assets, profits (without duplication) of group of agents equals total credits, assets, profits of the group of agents. Sum of agents additive economic and financial variables determine macroeconomic variables. Sum of agents investment, taxes, assets (without duplication) of all agents in economy equals macro investment, taxes, assets and etc. Agents also have non-additive variables. For example price, tax rates, bank rates are non-additive variables. Non-additive variables are determined as ratio of additive: price is the ratio of value to volume of market trades, bank rate is the ratio of cost of credit per time term to loan body and etc. Additive variables determine all non-additive variables. Hence description of economic processes and the business cycle in particular can use additive economic variables only. Almost all additive variables change due to market transactions executed by agents. Agents sell and buy commodities, assets, provide credits and investment and etc. These transactions change the amount of agents assets, commodities, investment and etc. Agents perform market transactions under action of their expectations. Agents create their expectations on base of their forecasts of trends of economic and financial variables, market activity and price dynamics, inflation and bank rates and etc. Agents expectations may reflect expectations of other agents, their spirits or weather forecasts and any other factors that can impact agents decisions go into particular market transaction. Agents expectations are the main tool that deliver influence and perturbations of social, mental, political, economical and any other factors on market transactions and macroeconomic evolution.

We regard methods and models that describe relations between economic variables, market transactions and expectations as the essence of theoretical economics.

*2.1. Risk assessment*

For decades, international risk rating agencies like Fitch, Moody's, S&P (Metz and Cantor, 2007; Chane-Kon, et.al, 2010; Kraemer and Vazza, 2012) assess credit risk ratings for largest banks and multinational corporations. Regular assessments of risk rating of major banks and corporations are reasons for taking investment decisions by largest financial institutions, pension funds and etc. Risk ratings take values of risk grades and are denoted by letters as *AAA, AA, BB* and etc. (Metz and Cantor, 2007; Chane-Kon, et.al, 2010, Kraemer and Vazza, 2012). Due to their economic activity and due to evolution of economic environment agents change their risk ratings during some time. Rating agencies evaluate and publish risk transition matrices that describe the probability that agent's risk may change from one risk grade to another during certain time term T (Belkin, 1998; Bangia, Diebold and Schuermann, 2000; Ho et.al, 2017; S&P, 2018). Each rating company uses its own letter notations of risk grades. Particular choice of risk grades is determined by risk methodology and rating companies use their own grades to protect and enlarge their own business. However at least 80 years ago Durand (1941) and then Myers and Forgy (1963) proposed numerical risk grades. Last years numerical risk grades are used as a tool to compare the risk assessments made by different rating agencies, as a method for development of the credit scoring and etc. (Becker and Milbourn, 2010; Morkoetter, Stebler and Westerfeld, 2016; King, Ongena and Tarashev, 2017; WB Group, 2019). We outline numerical risk grades because they bring the new insights into methods and models of theoretical economics. The choice between letters or numerical risk grades is determined by methodology and preferences of rating agencies. Any set of letter risk grades can be denoted as set of discrete points in space R. Such transition doesn't add much sense to risk assessment but helps simplify the comparison of risk assessments made by different rating agencies. We suggest take the next step and introduce continuous numerical risk grades. Usage of continuous risk grades substantially changes the role and the meaning of agents risk assessment. We don't consider here any particular methodology of continuous risk grades assessment for but explain the benefits of continuous risk grades for economic theory.

*2.2 Risk motion*

Many risks impact economic evolution. We don't consider particular risks and propose that one can chose 1, 2, 3 risks that impact economic agents and economic processes. For convenience we describe economy under action of $n$ risks, $n=1,2,..$ The absolute values of continuous risk grades are subject of the convenience of use. We chose continuous risk grades $x$ that fill the interval $[0,1]$ of space $R$. We take the 0 as the most secure and 1 as the most risky grades. Thus assessment of agents risk $x \in [0,1]$ for the single risk distribute all agents over interval [0,1]. Assessments of agents ratings for $n$ risks distribute agents over unit cube in $R^n$. Let's assume that at moment $t$ there are $N(t)$ agents in the economy and assessments of $n$ risks for all agents $i=1,...N(t)$ define risk ratings $\boldsymbol{x}_i$ of agent $i$ (2.1):

$$\boldsymbol{x}_i = (x_{1i}, \dots x_{ni}); \quad 0 \le x_{ji} \le 1 \; ; \quad i = 1,..N(t); \quad j = 1, \dots n \qquad (2.1)$$

Risk ratings $\boldsymbol{x}_i$ (2.1) of agent $i$ play role of coordinates in economic domain (2.2):

$$\boldsymbol{x} = (x_1, \dots x_n) \; ; \quad 0 \le x_j \le 1 \; ; \quad j = 1, \dots n \; ; \quad \boldsymbol{x} \in R^n \qquad (2.2)$$

For decades major rating agencies present distribution of largest banks and corporations by their risk ratings (Bangia, Diebold and Schuermann, 2000; Pompliano and Hancock, 2002; Volland, et.al. 2019). We suggest disseminate this practice on all economic agents in the economy. Actually it is common economic practice now. Rating agencies distribute banks by credit ratings and each bank provides credit ratings of its clients. Almost all agents are clients of banks now and hence credit ratings now are already assessed for almost all agents. We only suggest use the unified methodology for risk assessment and make current practice to assess bank client ratings publicly available. Distribution of economic agents by their risk ratings $\boldsymbol{x}$ as coordinates in economic domain (2.2) establishes distributions of their economic and financial variables by the risk coordinates in economic domain (Appendix A). Moreover, distribution of agents by their continuous numerical risk ratings $\boldsymbol{x}$ in economic domain (2.2), introduces velocities $\boldsymbol{v}$ of agents motion in domain (2.2). Motion of agents in the space of risk ratings introduces hidden and important properties of economic processes that impact the evolution of economic state and determine the nature of business and credit cycles in particular. Indeed, any economic activity of agents itself and various economic and financial processes, technological or political trends, any factors that impact economic development induce change of agent risk ratings. As we mentioned above, rating agencies evaluate risk transition matrices (Metz and Cantor, 2007; Moody's, 2009; Fitch, 2017; S&P, 2018) for largest banks and corporations. Elements $a_{ij}$ of transition matrices describe the probability $a_{ij}$ that agent with risk rating $x_i$ will move to rating $x_j$ during certain time term T. As usual interval $T$ equals one, two or three years. At present the risk ratings are denoted by letters. We suggest replace letter risk grades by continuous numerical risk grades. This replacement permit determine the distance between risk rating $\boldsymbol{x}_i$ and $\boldsymbol{x}_j$. Indeed, for numerical continuous grades the transition from rating $\boldsymbol{x}_i$ to $\boldsymbol{x}_j$ defines interval $\boldsymbol{l}_{ij}$ :

$$\boldsymbol{l}_{ij} = \boldsymbol{x}_j - \boldsymbol{x}_i \tag{2.3}$$

Transition from $\boldsymbol{x}_i$ to $\boldsymbol{x}_j$ takes time T and hence defines the velocity $\boldsymbol{v}_{ij}$ of the agents motion from $\boldsymbol{x}_i$ to $\boldsymbol{x}_j$ during time T as:

$$\boldsymbol{v}_{ij} = \frac{\boldsymbol{l}_{ij}}{T} \tag{2.4}$$

Element $a_{ij}$ of transition matrices define probability $a_{ij}$ of motion from $\boldsymbol{x}_i$ to $\boldsymbol{x}_j$ with velocity $\boldsymbol{v}_{ij}$ during time T. Hence transition matrix defines mean velocity $\boldsymbol{v}(t,\boldsymbol{x}_i)$ of agent at point $\boldsymbol{x}_i$ as:

$$\boldsymbol{v}(t,\boldsymbol{x}_i) = \sum_{j=1}^{K} \boldsymbol{v}_{ij} a_{ij} = \frac{1}{T}\sum_{j=1}^{K} \boldsymbol{l}_{ij} a_{ij} \;\; ; \;\; \sum_{j=1}^{K} a_{ij} = 1 \tag{2.5}$$

Here $K$ means the number of different numerical risk grades that defines the degree $K$x$K$ of the transition matrix. Motion of agent $i$ in economic domain (2.2) causes that agent $i$ caries its economic and financial variables with velocity $\boldsymbol{v}_i$. Collective effect generated by transport of economic and financial variables by individual agents during their risk motion in economic domain (2.2) establishes important and influential factor that impact macroeconomic evolution. Let’s consider this in more details.

*2.3 Collective economic behavior*

Economics is a social phenomenon. Description of economic processes requires description of collective economic behavior of individual agents. Agents risk assessment and distribution of individual agents by their risk ratings as coordinates in economic domain (2.2) establish the way for description of agents collective economic behavior and help develop continuous economic media approximation. The main reason for such approximation is the transition from description of economic and financial variables as properties of individual agents to description of distributions of economic and financial variables as functions of risk coordinates in economic domain (2.2). To evaluate such a transition one should rougher the description of agents coordinates. In simple words, if one has a meter ruler with divisions of millimeters, than to rougher the description one should aggregate millimeter scales in each centimeter division of the ruler and leave only centimeter divisions at the ruler. The new ruler measures the distance with less accuracy. Let’s do the similar and rougher description of agents in economic domain (2.2). For *n* risks and *n*-dimensional economic domain (2.2) let's chose a scale *d* such as:

$$0 < d < 1 \quad ; \quad dV(\boldsymbol{x}) \sim d^n \tag{2.6}$$

Let's assume that at point $\boldsymbol{x}$ a small volume $dV(\boldsymbol{x})$ contains a lot of agents $i$ with coordinates $\boldsymbol{x}_i$ inside the volume $dV(\boldsymbol{x})$. Let's take additive economic or financial variable $A_i(t,\boldsymbol{x})$ of agent $i$ inside the volume $dV(\boldsymbol{x})$ with risk coordinates $\boldsymbol{x}$ (2.1) and risk velocity $\boldsymbol{v}_i(t,\boldsymbol{x})$ (2.5). As variable $A$ one can consider agents assets, investment, credits, profits and etc. Let's sum values of variable $A_i(t,\boldsymbol{x})$ of all agents $i$ with risk coordinates $\boldsymbol{x}$ inside volume $dV(\boldsymbol{x})$ (2.6). Actually all economic assessments, measurements and observations take certain time term $\Delta$. As $\Delta$ one can consider minute, hour, week, month and etc. Time interval $\Delta$ determine internal time scale of the continuous economic approximation – all variations of economic variables or processes described by this particular approximation take time t> $\Delta$. Let's chose such time term $\Delta$ and averaged sum of agents variable $A_i(t,\boldsymbol{x})$ inside the volume $dV(\boldsymbol{x})$ during the interval $\Delta$ (A.1, A.2). We call the sum of values of variable $A_i(t,\boldsymbol{x})$ of all agents $i$ with risk coordinates $\boldsymbol{x}$ inside small volume $dV(\boldsymbol{x})$ (2.6) averaged during the interval $\Delta$ as the collective variable $A(t,\boldsymbol{x})$ (2.7):

$$A(t,\boldsymbol{x}) = \sum_{i \in dV(x);\, \Delta} A_i(t,\boldsymbol{x}) = \frac{1}{\Delta}\int_{t-\Delta/2}^{t+\Delta/2} d\tau \, \sum_{i \in dV(x)} A_i(\tau,\boldsymbol{x}) \tag{2.7}$$

Collective variable $A(t,\boldsymbol{x})$ is a function of $(t,\boldsymbol{x})$ and describes economic or financial properties of point $\boldsymbol{x}$ in the domain (2.2) but not the properties of individual economic agent. The same procedure defines other additive economic variables. It is obvious, that integral by $d\boldsymbol{x}$ over economic domain (2.2) of variable $A(t,\boldsymbol{x})$ or choice of sum in (2.7) by all agents in the entire economy defines macroeconomic variable $A(t)$ (A.3) as function of time $t$ averaged by interval $\Delta$.

$$A(t) = \int d\boldsymbol{x}\, A(t,\boldsymbol{x}) = \frac{1}{\Delta}\int_{t-\Delta/2}^{t+\Delta/2} d\tau \, \sum_{i \in E} A_i(\tau,\boldsymbol{x}) \tag{2.8}$$

Definition (2.8) matches the standard notion of macroeconomic variables $A(t)$ as sum of variables $A_i(t,\boldsymbol{x})$ of all agents $i$ in the economy (*E*) averaged during time term $\Delta$. Definition (2.7) of economic variable $A(t,\boldsymbol{x})$ gives intermediate approximation between description of variable $A_i(t,\boldsymbol{x})$ of individual agent $i$ and description of macroeconomic variable $A(t)$ (2.8) as sum over all agents in the economy. Such intermediate approximation uncovers hidden processes that impact economic evolution and are almost neglected by current macroeconomic models.

The hidden factors that impact macroeconomic evolution are the collective flows of economic or financial variables induced by motion of economic agents in economic domain (2.2). To explain our statement let's chose variable $A_i(t,\boldsymbol{x})$ of agent $i$ at point $\boldsymbol{x}_i$ at moment $t$. Agent $i$ moves in economic domain (2.2) with velocity $\boldsymbol{v}_i(t,\boldsymbol{x})$ (2.5). It is said that product $A_i(t,\boldsymbol{x})\boldsymbol{v}_i(t,\boldsymbol{x})$ describes the flow $\boldsymbol{p}_{iA}(t,\boldsymbol{x})$ (A.4) of the variable $A_i(t,\boldsymbol{x})$ that agent $i$ carries with velocity $\boldsymbol{v}_i(t,\boldsymbol{x})$. Different agents $i$ carries different flows $\boldsymbol{p}_{iA}(t,\boldsymbol{x})$ of variable $A_i(t,\boldsymbol{x})$ with different velocities $\boldsymbol{v}_i(t,\boldsymbol{x})$ (2.5). Collective transport of economic variable $A_i(t,\boldsymbol{x})$ by all agents inside volume $dV(\boldsymbol{x})$ averaged during interval $\Delta$ determines the collective flow $\boldsymbol{P}_A(t,\boldsymbol{x})$ (A.5):

$$\boldsymbol{P}_A(t,\boldsymbol{x}) = \sum_{i\in dV(\boldsymbol{x});\,\Delta} A_i(t,\boldsymbol{x})\boldsymbol{v}_i(t,\boldsymbol{x}) = A(t,\boldsymbol{x})\boldsymbol{v}_A(t,\boldsymbol{x}) \tag{2.9}$$

of the collective variable $A(t,\boldsymbol{x})$ (2.7) with the collective velocity $\boldsymbol{v}_A(t,\boldsymbol{x})$ (2.9). Different economic variables $A(t,\boldsymbol{x})$, $B(t,\boldsymbol{x})$, $C(t,\boldsymbol{x})$ define different flows $\boldsymbol{P}_A(t,\boldsymbol{x})$, $\boldsymbol{P}_B(t,\boldsymbol{x})$, $\boldsymbol{P}_C(t,\boldsymbol{x})$ (2.9) with different velocities $\boldsymbol{v}_A(t,\boldsymbol{x})$, $\boldsymbol{v}_B(t,\boldsymbol{x})$, $\boldsymbol{v}_C(t,\boldsymbol{x})$. Relations and interactions between variables, flows and their velocities establish harsh environment for macroeconomic modeling. Relations (2.10) cause that macroeconomic variable $A(t)$ (2.8) moves in the economic domain with the velocity $\boldsymbol{v}_A(t)$. Indeed, if one takes integral (A.6) by $\boldsymbol{P}_A(t,\boldsymbol{x})$ (2.9) over economic domain (2.2) or sum the product $A_i(t,\boldsymbol{x})\boldsymbol{v}_i(t,\boldsymbol{x})$ by all agents $i$ in the entire economy $E$ than one derives macroeconomic flow $\boldsymbol{P}_A(t)$ (2.10) of economic variable $A(t)$ (2.8):

$$\boldsymbol{P}_A(t) = A(t)\boldsymbol{v}_A(t) = \int d\boldsymbol{x}\, A(t,\boldsymbol{x})\boldsymbol{v}_A(t,\boldsymbol{x}) \tag{2.10}$$

Relations (2.10) determine very important notion: the motion of macroeconomic variable $A(t)$ with the velocity $\boldsymbol{v}_A(t)$ in economic domain (2.2). Velocity $\boldsymbol{v}_A(t)$ cause change of new important economic notion $\boldsymbol{v}_A(t)$ - the mean risk $\boldsymbol{X}_A(t)$ (A.7) of economic variable $A(t,\boldsymbol{x})$. We introduce the mean risk $\boldsymbol{X}_A(t)$ (A.7) of economic variable $A$ as sum of products $\boldsymbol{x}_i A(t,\boldsymbol{x}_i)$ for all agents in the economy (A.7) weighed by variable $A(t)$ and averaged during interval $\Delta$:

$$\boldsymbol{X}_A(t)A(t) = \int d\boldsymbol{x}\,\boldsymbol{x}\,A(t,\boldsymbol{x}) = \frac{1}{\Delta}\int_{t-\Delta/2}^{t+\Delta/2} d\tau\ \sum_{i\in E} \boldsymbol{x}\ A_i(\tau,\boldsymbol{x}) \tag{2.11}$$

Risk assessment of economic agent $i$ determines its risk coordinates $\boldsymbol{x}$ in economic domain (2.2). The notion $\boldsymbol{X}_A(t)$ (A.7) solves the problem of mean risk associated with economic variable $A$. Different economic variables have different impact on economic evolution. To assess risks of entire economy one should assess risks related with economic variables. Continuous numerical risk grades open the way for introducing the mean risks of economic variables. As we show below dynamics of mean risks of economic variables in economic domain (2.1, 2.2) describes phases of business cycles. Econometrics and observation of mean risks may help reduce financial crises losses and develop more sustainable economic policy. Relations (2.11) define the mean risk $X_A(t)$ of related with economic variable $A(t)$ (2.8) that is somewhat similar to VWAP (Berkowitz et.al 1988; Buryak and Guo, 2014; CME Group, 2020) – it is average risk weighted by variable $A(t,\boldsymbol{x}_i)$. We emphasize that different economic variables define different mean risks. Mean risk $X_I(t)$ of investment $I(t)$ in the economy is different from mean risk $X_C(t)$ of credits $C(t)$, or mean risk $X_D(t)$ of demand $D(t)$, or mean risk $X_{VA}(t)$ of value added $VA(t)$ and etc. Agents economic activity, regulatory, political, technology variations change agents risk coordinates and hence change mean risks of macroeconomic and financial variables. Below we present a simple model that describe the motion of variable $A(t)$ with velocity $\boldsymbol{v}_A(t)$ in economic domain (2.2) and corresponding motion of the mean risk $X_A(t)$ with different velocity. (App.C.). Borders of economic domain (2.2) of a unit cube in $R^n$ limit change of the mean risks of any economic variable. Hence motion of mean risk $X_A(t)$ of any economic variable $A(t)$ should fluctuate from secure to risky areas of economic domain (2.2) and back. These slow oscillations of mean risks of macroeconomic variables reflect slow change of mean risks of economic variables in the economy. Interrelations between different economic and financial variables cause correlations between motion of different mean risks and establish a complex picture of macroeconomic evolution. Economics is a complex system with strong backward interactions. The change of mean risks $X_A(t)$ impact change of value of macro variable $A(t)$ or vice versa. Thus oscillations of the mean investment risk from secure to risky area in economic domain (2.2) and back can describe the macroeconomic investment cycles from low to high investment activity and then back to low. Credit cycles follow the fluctuations of mean credit risks and etc. Growth of economic activity and rise of demand should be accompanied with motion of demand mean risk from secure to risky area in (2.2). We state that cyclical motions of mean risks of economic variables in economic domain (2.2) reflect essence of business and credit cycles and express the nature of economic and financial crises. Variety of factors those impact motion of mean risks, mutual interactions between macroeconomic variables and their flows, other reasons that define properties of oscillations of mean risks in economic domain (2.2) establish a complex dynamics of business cycles and economic crises. Each new one will be different from the previous but all such events are accompanied by cycles of mean risk in economic domain.

In App.C we present a simple model of business cycle. In the next Section we study processes that determine change of macroeconomic and financial variables.

## 3. Collective transactions

The conventional economic models describe mutual dependence of macroeconomic variables. For example the change of demand causes the change of supply, variations of cost of capital causes the change investment activity and etc. Of course, these economic “laws” are correct, but they reflect indirect relations between macroeconomic variables. The only processes those have direct effects on change of economic and financial variables are described by market deals, trades, transactions between economic agents. Only trading goods, commodities, currencies, assets, capital, funds, service and etc., change agents economic and financial variables and hence impact the change of macroeconomic variables. Any variation of market regulation, tax policy, custom decisions, the spread of information on rise or fall of demand or supply, financial markets or national currency, and etc., take economic effect if and only after the deals, trades, and transactions between agents are executed. Only market trading impacts economic growth, price trends, cause business cycles or approaching crisis. Deals and trades are the only direct drivers of the economic development. All other economic factors, economic trends, financial policy, market information, forecasts and etc., forms agents expectations and these expectations impact agents decisions to take or reject trades, deals and transactions. For sure agents expectations take effect on trade decisions but only executed transactions establish records for the change of economic variables. Description of trades is the only way to understand the hidden rules, reasons and effects of economic processes.

To describe trades between agents and model collective impact of transactions on economic processes let's take into account the same factors we use to describe collective impact of agents economic and financial variables. Agents perform transactions with different assets, commodities, currencies and etc. Let's chose trades with particular commodities or assets and study only this type of transactions. Let's assume that agent $i$ at point $\boldsymbol{x}$ sells the volume $U_{ij}$ of selected assets, commodities or service, etc. that we denote as variable $A$ to agent $j$ at point $\boldsymbol{y}$ and agent $j$ pays the value $C_{ij}$ for the for the volume $U_{ij}$ of selected variable $A$. As variable $A$ one can consider securities, commodities, gold, crude oil, any specific assets, investment and etc. We define the buy-sell transaction $\boldsymbol{bs}_{ij}$ between agents $i$ and $j$ as two-component function (3.1):

$$\boldsymbol{bs}_{ij}(t,\boldsymbol{z}) = \left(U_{ij}(t,\boldsymbol{z}); C_{ij}(t,\boldsymbol{z})\right) \; ; \; \boldsymbol{z} = (\boldsymbol{x},\boldsymbol{y}) \tag{3.1}$$

Risk coordinates $\boldsymbol{x}$ and $\boldsymbol{y}$ of agents $i$ and $j$ involved into trade $\boldsymbol{bs}_{ij}$ (3.1) define economic domain (3.2; 3.3) as unit cube with dimension $2n$ in $R^{2n}$:

$$\boldsymbol{z} = (\boldsymbol{x},\boldsymbol{y}) \;\; ; \;\; \boldsymbol{x} = (x_1 \dots x_n) \;\; ; \;\; \boldsymbol{y} = (y_1 \dots y_n) \tag{3.2}$$

$$0 \le x_i \le 1 \;\; ; 0 \le y_j \le 1 \; ; \;\; i = 1,..n \; ; \; j = 1,..n \tag{3.3}$$

The volume $U_{ij}$ of the trade $\boldsymbol{bs}_{ij}$ (3.1) change the amount of variable $A$ of agents $i$ and $j$ involved into the trade and the value $C_{ij}$ change the amount of funds of agents $i$ and $j$. In real economy the transaction $\boldsymbol{bs}_{ij}$ (3.1) with volume $U_{ij}$ of variable $A$ and payment of the value $C_{ij}$ can be executed in different moments but for simplicity we assume that the deals $\boldsymbol{bs}_{ij}$ (3.1) with the volume $U_{ij}$ and the value $C_{ij}$ are performed simultaneously. Precise description of transactions between individual agents is almost impossible. Indeed, any attempt to collect exact information about transactions of all agents simply can destroy the market and disturb economic environment. Thus one should replace exact description of the transactions $\boldsymbol{bs}_{ij}$ (3.1) between individual agents $i$ and $j$ by description of collective trades of agents alike to description of collective economic variables $A(t,\boldsymbol{x})$ (2.7). To do that we take (2.6) and introduce small volume $dV(z)$ in economic domain (3.2; 3.3):

$$dV(\boldsymbol{z}) = dV(\boldsymbol{x})dV(\boldsymbol{y}) \ ; \ dV(\boldsymbol{x}) \sim d^n \quad ; \quad dV(\boldsymbol{y}) \sim d^n \tag{3.4}$$

We assume that volume $dV(\boldsymbol{x})$ contain many sellers and volume $dV(\boldsymbol{y})$ contain many buyers involved into trade $\boldsymbol{bs}_{ij}$ (3.1) with variable $A$. We aggregate the volumes $U_{ij}$ and the values $C_{ij}$ of all trades between sellers inside $dV(\boldsymbol{x})$ and all buyers inside $dV(\boldsymbol{y})$ and average it during time term $\Delta$. Thus we replace description of trades $\boldsymbol{bs}_{ij}$ (3.1) between individual agents $i$ and $j$ by description of collective transactions $\boldsymbol{BS}(t,z)$ (B.2-B.4) at point $z=(\boldsymbol{x},\boldsymbol{y})$ of economic domain (3.2; 3.3).

$$\boldsymbol{BS}(t,\boldsymbol{z}) = \big(U(t,\boldsymbol{z}); C(t,\boldsymbol{z})\big) \quad ; \quad \boldsymbol{z} = (\boldsymbol{x},\boldsymbol{y}) \tag{3.5}$$

$U(t,z)$ and $C(t,z)$ describe the volume and value of the transactions at point $z=(\boldsymbol{x},\boldsymbol{y})$ between agents inside $dV(\boldsymbol{x})$ and $dV(\boldsymbol{y})$ averaged during interval $\Delta$. As we mentioned above, economic activity of agents and variations of economic environment cause change of agents risk coordinates. Seller $i$ and buyer $j$ involved into transaction $\boldsymbol{b}_{ij}$ (3.1) move in economic domain and this motion cause flows $\boldsymbol{p}_{ijU}$ of the volume $U_{ij}$ and flows $\boldsymbol{p}_{ijC}$ of the value $C_{ij}$ of the trade $\boldsymbol{b}_{ij}$ (B.5-B.8). Collective flows of the volume $U_{ij}$ and the value $C_{ij}$ between all agents inside volume $dV(\boldsymbol{x})$ and volume $dV(\boldsymbol{y})$ and averaged during interval $\Delta$ defines the flow $\boldsymbol{P}(t,z)$ (B.9-B.17) of the collective transaction $\boldsymbol{BS}(t,z)$. Flow $\boldsymbol{P}(t,z)$ describes the transport of the collective volume $U(t,z)$ and collective value $C(t,z)$ of the trade $\boldsymbol{BS}(t,z)$ with velocity $\boldsymbol{v}(t,z)$.

$$\boldsymbol{P}(t,\boldsymbol{z}) = \big(\boldsymbol{P}_U(t,\boldsymbol{z}), \boldsymbol{P}_C(t,\boldsymbol{z})\big) \ ; \ \boldsymbol{z} = (\boldsymbol{x},\boldsymbol{y}) \tag{3.6}$$

$$\boldsymbol{P}_U(t,\boldsymbol{z}) = U(t,\boldsymbol{z})\boldsymbol{v}_U(t,\boldsymbol{z}) \quad ; \quad \boldsymbol{P}_C(t,\boldsymbol{z}) = C(t,\boldsymbol{z})\boldsymbol{v}_C(t,\boldsymbol{z}) \tag{3.7}$$

Relations (3.7) describe flow $\boldsymbol{P}_U(t,z)$ of the volume $U(t,z)$ and flow $\boldsymbol{P}_C(t,z)$ of the value $C(t,z)$ of transaction (3.5). Relations (B.12) determine flow $\boldsymbol{P}_{xU}(t,z)$ of the volume $U(t,z)$ along axis X with velocity $\boldsymbol{v}_{xU}(t,z)$ that is induced by motion of the sellers at point $\boldsymbol{x}$. Relations (B.13) determine flow $\boldsymbol{P}_{yU}(t,z)$ of the value $U(t,z)$ along axis Y with velocity $\boldsymbol{v}_{yU}(t,z)$ that is induced by motion of the buyers at point $\boldsymbol{y}$. The similar relations define flows of the value of the transaction $\boldsymbol{B}(t,z)$ (B.14, B.15). Trades with different assets, commodities or currencies have different flows and different velocities in economic domain (3.2; 3.3). Flows and velocities of trades with crude oil are different from flows and velocities of trades with gold, cupper, wheat and etc. Aggregations of transactions performed between sellers and buyers in the entire economy define the collective macroeconomic transaction $\boldsymbol{BS(t)}$, its flow $\boldsymbol{P(t)}$ and velocity $\boldsymbol{v(t)}$ in domain (3.2; 3.3). Similar to motion of macroeconomic variable $A(t)$ with velocity $\boldsymbol{v}_A(t)$ in economic domain (2.2) flow $\boldsymbol{P(t)}$ and velocity $\boldsymbol{v(t)}$ describe collective transport of the macro transaction $\boldsymbol{BS(t)}$ induced by risk motion of sellers and buyers (B.21-B.32):

$$\boldsymbol{BS}(t) = \big(U(t); C(t)\big) \quad ; \quad \boldsymbol{P}(t) = \big(\boldsymbol{P}_U(t); \boldsymbol{P}_C(t)\big) \tag{3.8}$$

$$\boldsymbol{P}_U(t) = U(t)\boldsymbol{v}_U(t) \quad ; \quad \boldsymbol{P}_C(t) = C(t)\boldsymbol{v}_C(t) \tag{3.9}$$

Relations (3.8) define macro transaction $\boldsymbol{BS(t)}$ (B.21) as pair of the total volume $U(t)$ (B.22) and the total value $C(t)$ (B.22) of all transactions $\boldsymbol{BS(t)}$ in the economy averaged during interval $\Delta$. It is important to emphasize that risk motions of sellers involved into transaction $\boldsymbol{BS(t)}$ induce macro flows $\boldsymbol{P}_{xU}(t)$ (B.27) of the volume $U(t)$ along axis $X$ and risk motions of buyers induce flows $\boldsymbol{P}_{yU}(t)$ (B.28) of the volume $U(t)$ along axis $Y$ in economic domain. The similar flows $\boldsymbol{P}_{xC}(t)$ and $\boldsymbol{P}_{yC}(t)$ (B.29, B.30) carry the total value $C(t)$. Flows of the volume $\boldsymbol{P}_{xU}(t)$, $\boldsymbol{P}_{yU}(t)$ and the value $\boldsymbol{P}_{xC}(t)$ and $\boldsymbol{P}_{yC}(t)$ have different velocities and the agents motion is limited by borders of economic domain (3.2, 3.3). Thus velocities of the volume $\boldsymbol{v}_{xU}(t)$, $\boldsymbol{v}_{yU}(t)$ and the value $\boldsymbol{v}_{xC}(t)$ and $\boldsymbol{v}_{yC}(t)$ (B.32) should fluctuate and change direction from secure to risky area of the domain (3.2, 3.3) and back. Motions of the volume $U(t)$ and the value $C(t)$ of transactions $\boldsymbol{BS(t)}$ (3.8) have similar nature as motions of economic variables. Similar to mean risk $\boldsymbol{X}_A(t)$ of variable $A(t)$ (2.11) transactions $\boldsymbol{BS(t)}$ (3.5, 3.8) define mean risk $\boldsymbol{X}_U(t)$ of the volume $U(t)$ and mean risk $\boldsymbol{X}_C(t)$ of the value $C(t)$ (B.33-B.36). We introduce mean risk $\boldsymbol{X}_{BS}(t)$ of transactions $\boldsymbol{BS(t)}$ as two component function (B.33-B.36):

$$\boldsymbol{X}_{BS}(t) = \big(\boldsymbol{X}_U(t); \boldsymbol{X}_C(t)\big) \tag{3.10}$$

Evolution of volume $U(t,z)$ and value $C(t,z)$ (B.2) and dynamics of flows of volume $\boldsymbol{P}_U(t,z)$ and value $\boldsymbol{P}_U(t,z)$ (B.9) cause change of $U(t)$ and $C(t)$ (B.21) and $\boldsymbol{P}_U(t)$ and $\boldsymbol{P}_U(t)$ (B.23) and generate motion of mean risks $\boldsymbol{X}_U(t)$ and $\boldsymbol{X}_C(t)$ in economic domain (3.2, 3.3). Economic domain (3.2, 3.3) is a unit cube and hence motions of mean risks $\boldsymbol{X}_U(t)$ and $\boldsymbol{X}_C(t)$ present complex fluctuations from secure to risky area of economic domain and back. We regard fluctuations of mean risks in economic domain as properties of business cycle. Any business cycle is accompanied by fluctuations of the mean risks of collective transactions.

## 4. Fluctuations of the mean risks – the nature of business cycles

Economics is a complex mixture of numerous processes with direct and backward interactions and constraints. The objective of theoretical economics is the development of successive approximations of real economic processes. Any approximation simplifies the economic reality. We chose the risk assessment as the tool for distributing agents by economic domain and describe the motion of economic variables and market transactions as definite approximation of real economic processes. We regard the fluctuations of the mean risks $X_A(t)$ of economic variable $A(t)$ as the core essence of cycle evolution. The cycles of the credits and investment, demand and production function are accompanied by and described by fluctuations of the corresponding mean risks. The mean risk $X_A(t)$ of variable $A(t)$ change its value from secure to risky area of economic domain and back. Along with the motion of mean risk the corresponding variable $A(t)$ changes from growth to stagnation, then decline and so on. It is common knowledge that in secure state agents economic activity is low, but agents preferences and desires move them take the risky decisions under risky expectations and execute the risky transactions. That increases agents economic activity but moves them to more risky area of economic domain. Different agents act differently in same risk conditions. Aggregation of their collective economic activity, aggregation of agents variables and transactions smooth description of the individual agents dynamics and presents description of collective economic behavior. Description of collective agents activity establishes collective flows of agents variables and transactions. Flows of variables and transactions determine fluctuations of mean risks of collective motion in economic domain. The factors that govern the mean risk fluctuations in economic domain, the exogenous or endogenous "shocks" that initiate the motion of mean risks from secure to risky area of economic domain or back can be different. Each new business cycle differs from the previous one. We emphasize that any business cycle, any cyclical change of economic growth to stagnation and economic decline are accompanied by the mean risk motion. Description of mean risk motion gives approximation of the business cycles.

### *4.1. Risk motion equations*

Evolution of collective economic variables and transactions determine the mean risk motion. To reduce the complexity of the current paper we omit here description of collective expectations and refer to (Olkhov, 2019b; 2019c) for modeling the collective expectations, their flows and their impact on transactions. However we confirm that methods presented in Sec. 2 and 3 permit describe collective expectations and their impact on market transactions.

To simplify further description and introduce the self-consistent business cycle model we reduce our study by equations on collective economic variables $A(t,\boldsymbol{x})$ (2.7) (Olkhov, 2016; 2017; 2018a). As variable $A(t,\boldsymbol{x})$ (2.7) one can consider collective investment, credits, assets and etc., of agents with risk ratings near $\boldsymbol{x}$. Let's take arbitrary small volume $dV$ in economic domain (2.2). The change of variable $A(t,\boldsymbol{x})$ in a volume $dV$ during time $dt$ is determined by two factors. The first one describes the change in time of $A(t,\boldsymbol{x})$ during $dt$ in a volume $dV$ :

$$\int dV \, \frac{\partial}{\partial t} A(t,\boldsymbol{x}) \tag{4.1}$$

The second factor is determined by the flows $\boldsymbol{P}_A(t,\boldsymbol{x})=A(t,\boldsymbol{x})\boldsymbol{v}_A(t,\boldsymbol{x})$ (2.9) of agents that may flow in- or flow out- through the surface of small volume $dV$ during time $dt$. Agents that flow in- a volume $dV$ during $dt$ increase the amount of variable $A$ in a volume $dV$. Agents that flow out of the surface of volume $dV$ decrease the amount of variable $A$ inside volume $dV$. Total change of variable $A$ in volume $dV$ due to in- and out- flows $\boldsymbol{P}_A(t,\boldsymbol{x})$ (2.9) takes form of surface integral:

$$\oint ds\, \boldsymbol{P}_A(t,\boldsymbol{x}) = \oint ds\, A(t,\boldsymbol{x})\, \boldsymbol{v}_A(t,\boldsymbol{x}) \tag{4.2}$$

Integral in (4.2) is taken by the surface of volume $dV$. The well-known Gauss' Theorem (Strauss 2008, p.179) states that the surface integral by the flow equals the volume integral by divergence of the flow and thus (4.2) takes form:

$$\oint ds \;\; A(t,\boldsymbol{x})\boldsymbol{v}_A(t,\boldsymbol{x}) = \int dV \;\; \nabla\cdot\big(A(t,\boldsymbol{x})\boldsymbol{v}_A(t,\boldsymbol{x})\big) \tag{4.3}$$

Hence total change of variable $A$ in arbitrary small volume $dV$ equals sum of (4.1) and (4.3):

$$\int dV \left[\frac{\partial}{\partial t} A(t,\boldsymbol{x}) + \nabla\cdot\big(A(t,\boldsymbol{x})\boldsymbol{v}_A(t,\boldsymbol{x})\big)\right] \tag{4.4}$$

As volume $dV$ is arbitrary small hence the change of variable $A(t,\boldsymbol{x})$ at point $\boldsymbol{x}$ takes the form:

$$\frac{\partial}{\partial t} A(t,\boldsymbol{x}) + \nabla\cdot\big(A(t,\boldsymbol{x})\boldsymbol{v}_A(t,\boldsymbol{x})\big) = F_A(t,\boldsymbol{x}) \tag{4.5}$$

Function $F_A(t,\boldsymbol{x})$ in (4.5) describes any other factors that impact the change of the variable $A(t,\boldsymbol{x})$. Equation (4.5) use the flow $\boldsymbol{P}_A(t,\boldsymbol{x})=A(t,\boldsymbol{x})\boldsymbol{v}_A(t,\boldsymbol{x})$ (2.9) of variable $A(t,\boldsymbol{x})$ (2.7). To derive equations on the flow $\boldsymbol{P}_A(t,\boldsymbol{x})$ one should repeat above reasoning with respect to the flow $\boldsymbol{P}_A(t,\boldsymbol{x})$. Equations of the flow $\boldsymbol{P}_A(t,\boldsymbol{x})$ (2.9) of variable $A(t,\boldsymbol{x})$ (2.7) takes the form (Olkhov, 2016; 2017; 2018a; 2018b)

$$\frac{\partial}{\partial t} \boldsymbol{P}_A(t,\boldsymbol{x}) + \nabla\cdot\big(\boldsymbol{P}_A(t,\boldsymbol{x})\boldsymbol{v}_A(t,\boldsymbol{x})\big) = \boldsymbol{F_P}(t,\boldsymbol{x}) \tag{4.6}$$

Function $\boldsymbol{F_P}(t,\boldsymbol{x})$ in the right side of (4.6) describes any factors that impact the change of the flow $\boldsymbol{P}_A(t,\boldsymbol{x})$ (2.9). Functions $F_A(t,\boldsymbol{x})$ and $\boldsymbol{F_P}(t,\boldsymbol{x})$ can describe action of other variables, transactions, expectations or their flows. We emphasize that flows of economic variables, transactions and expectations are induced by change of collective risk ratings of economic agents in the economic domain. These flows describe new factors that impact economic evolution and have not been considered in standard mainstream models before. Integral of equation (4.5) by $\boldsymbol{dx}$ over economic domain (2.2) gives ordinary differential equation on variable $A(t)$ (2.8):

$$\frac{d}{dt}A(t) = F_A(t)\,; \quad F_A(t) = \int d\boldsymbol{x}\, F_A(t,\boldsymbol{x}) \tag{4.7}$$

Equation (4.7) is a consequence of (4.3): there are no economic agents outside of domain (2.2) (Olkhov, 2016; 2017; 2018a; 2018b). Hence all flows outside (2.2) equal zero. Thus integral in (4.3) by any surface that include domain (2.2) equals zero. Simplicity of equation (4.7) hides complex dependence of function $F_A(t)$ on variables, transactions and expectations and their flows. Similar considerations permit derive equations on flows $\boldsymbol{P}_A(t)$ (2.10) of variable $A(t)$ (2.8).

$$\frac{d}{dt}\boldsymbol{P}_A(t) = \boldsymbol{F_P}(t)\,; \quad \boldsymbol{F_P}(t) = \int d\boldsymbol{x}\, \boldsymbol{F_P}(t,\boldsymbol{x}) \tag{4.8}$$

Functions $F_A(t)$ and $\boldsymbol{F}_P(t)$ in (4.7; 4.8) hide the complex relations described by (4.5, 4.6). Flows (2.9, 2.10) of variables and transactions and mean risks (2.11) and their functions were never taken into account as factors that impact evolution of macro variables $A(t)$ (4.7). Above consideration highlights the hidden complexities of economic and business cycle modeling. To avoid excess complexity in this paper we don't consider here equations on transactions and refer to (Olkhov, 2018b; 2019a; 2019b; 2019c) for further details.

*4.2. The model of the supply-demand cycle*

We present the simple model of supply $S(t)$ and demand $D(t)$ cycles accompanied by fluctuations of mean supply risk $X_s(t)$ and mean demand risk $X_d(t)$ in 1-dimensinal economic domain [0,1] that describes economic processes under action of a singe risk (App.C). As such a risk one may consider credit risk, inflation risk or any single risk that impact macroeconomics. We take into account interactions between supply and demand only and neglect impact of all other economic variables, transactions and expectations to make the model equations as simple as possible. Such simplification allows formulate self-consistent equations that describe mutual interaction between supply and demand and derive simple solutions that describe fluctuations of supply, demand and their mean risks. We consider supply-demand cycle in the linear approximation by disturbances of supply and demand. In the linear approximation by disturbances we derive equations that describe oscillations of supply and demand disturbances and equations that describe fluctuations of supply and demand mean risks. Even this toy model of supply and demand interactions emphasize relatively complex dynamics of supply and demand mean risks. Oscillations of supply $s(t)$ and demand $d(t)$ (C.15) disturbances are accompanied by oscillations of mean supply risk $x_s(t)$ (C.23) and mean demand risk $x_d(t)$ (C.24) disturbances with the same frequency $\omega$ (C.22) determined by oscillations of supply and demand velocities $v_s(t)$ and $v_d(t)$ (C.10-C.13).

$$s(t) = \frac{a}{\alpha} v_{sx0} \sin \omega t \quad ; \quad d(t) = -b \frac{S_0}{\omega D_0} v_{sx0} \cos \omega t \tag{4.9}$$

$$x_s(t) = \frac{a}{\alpha} \, v_{sx20} \sin \omega t - 2 \frac{v_{s0}}{\omega} \cos \omega t \tag{4.10}$$

$$x_d(t) = 2 \frac{S_0}{\alpha D_0} \, v_{s0} \sin \omega t - b \frac{S_0}{\omega D_0} (2 v_{sx20} - v_{sx0}) \cos \omega t \tag{4.11}$$

We refer App.C for further details.

## 5. Conclusion

Relations (4.9-4.11) describe cycles of the supply $s(t)$ and the demand $d(t)$ disturbances accompanied by fluctuations of the mean supply risk $x_s(t)$ and the mean demand risk $x_d(t)$ disturbances with the same frequency $\omega$. Time derivative of the supply $s(t)$ disturbances (C.9) depends upon factor $D_0 v_{dx}(t)$ that models the product of risk $x$ and flow of demand disturbances $D_0 v_d(t,x)$ integrated by economic domain. The similar product of risk $x$ and flow of supply disturbances $S_0 v_s(t,x)$ integrated by economic domain (C.10) impact time derivative of demand disturbances $d(t)$. Such factors that impact time derivatives of supply and demand were not studied before in business cycle models. One may probably object that the simplicity of the model assumptions doesn't fully reflect economic reality of the observed supply and demand cycles. It would be surprising if the real supply-demand cycle can be described by pure analytical expressions (4.9-4.11) without taking into account numerous factors that define impact of macroeconomic environment. Our consideration of the macroeconomic modeling and business cycles in particular demonstrate the complexity of the economic processes. A uniform approach to description of business cycles requires development of the successive approximations. We intentionally simplified the cycle model equations to underline the main advantages of our approach. We highlight that any business cycle that is driven by any reasons and that describes fluctuations of any macro variables is characterized by the mean risk fluctuations. Cycles of investment, economic growth, supply-demand cycle and etc., are characterized by fluctuations of corresponding mean risks. In our simple model the mean risk fluctuations are induced by fluctuations of supply and demand velocities. In reality the relations between fluctuations of flows, fluctuations of macro variables and their mean risks may be more complex, but flows play the core role in any business cycles. Nevertheless relations between risk ratings migration and the business cycle were studied at least for twenty years (Bangia, Diebold and Schuermann, 2000) our proposal for usage of continuous numeric risk grades gives a new look on meaning of ratings migration matrices and their role in business cycles.

In the conclusion we collect the main issues of our approach to description of the business cycle. The initial is the proposal to use continuous numerical risk grades as meters for risk ratings assessments. We chose that risk grades take value inside economic domain - a unit cube with most secure grade equals 0 and most risky equals 1. The second – assumption that risk assessments can be performed for almost all economic agents. If so, well-know risk transition matrices determine velocities of agents motion in economic space. To smooth irregular and complex picture of millions of economic agents identified by their risk coordinates, risk velocities and numerous economic and financial variables and market transactions between agents we introduce continuous economic approximation. We define collective variables, transactions and their flows aggregated by all agents inside small volume $dV$ in economic domain and averaged during time interval $\Delta$. We derive equations alike to equations of continuous media, those describe evolution of economic variables, transactions and their flows. Any flows inside a bounded economic domain (unit cube) should fluctuate from secure to risky area. These fluctuations of flows cause business cycles. Integrals by equations over economic domain give ordinary equations on time derivatives for collective variables, transactions and their flows as functions of time only. Such a long road to simple ordinary time derivatives discovers new economic factors – collective flows, velocities, mean risks and etc., those impact macroeconomic and business cycle dynamics. We present pure theoretical considerations. Econometric data that may support or reject our model doesn't exist. However we hope that our results will help develop risk assessment and econometrics to obtain reasonable and adequate approximations of complex economic processes and business cycles.

## Appendix A

### Economic variables, their flows and mean risks

We model economy as set of economic agents involved in numerous market transactions under action of $n$ risks. Economic agents are described by their risk ratings as coordinates $\boldsymbol{x}=(x_1,...x_n)$ on $n$-dimensional economic domain (2.1, 2.2). Agent $i$ with coordinates $(t,\boldsymbol{x})$ has additive variables $A_i(t,\boldsymbol{x})$ like investment, demand, profits and etc. The purpose of our approximation – move description of variables $A_i(t,\boldsymbol{x})$ as properties of individual agents to description of collective variables $A(t,\boldsymbol{x})$ as functions of coordinates $(t,\boldsymbol{x})$. To do that let's take certain time interval $\Delta$ and chose a unit volume $dV(\boldsymbol{x})$ (2.6). Let's collect variables $A_i(t,\boldsymbol{x})$ of agents inside $dV(\boldsymbol{x})$ (2.6) and average the sum during interval $\Delta$. We define variable $A(t,\boldsymbol{x})$ as:

$$A(t,\boldsymbol{x}) = \sum_{i\in dV(\boldsymbol{x});\,\Delta} A_i(t,\boldsymbol{x}) \tag{A.1}$$

$$\sum_{i\in dV(\boldsymbol{x});\,\Delta} A_i(t,\boldsymbol{x}) \equiv \frac{1}{\Delta}\int_{t-\Delta/2}^{t+\Delta/2} d\tau \, \sum_{i\in dV(\boldsymbol{x})} A_i(\tau,\boldsymbol{x}) \tag{A.2}$$

$i \in dV(\boldsymbol{x})$ denotes that agent $i$ belong to a unit volume $dV(\boldsymbol{x})$ (2.6). Time averaging smooth changes of $A(t,\boldsymbol{x})$. Integral by economic domain (2.1, 2.2) of $A(t,\boldsymbol{x})$ (A.1) defines $A(t)$ (A.3):

$$A(t) = \int d\boldsymbol{x}\; A(t,\boldsymbol{x}) \tag{A.3}$$

Relations (A.3) define the sum of variable $A(t,\boldsymbol{x})$ (A.1) over all agents in the economy averaged during interval $\Delta$. Agents economic activity and evolution of the economic environment cause change of agents risk ratings and define velocities $\boldsymbol{v}=(v_1,...v_n)$ of economic agents (2.3-2.5). Agent $i$ with velocity $\boldsymbol{v}_i$ carries variable $A_i(t,\boldsymbol{x})$ in economic domain (2.1, 2.2) and creates agent's flow $\boldsymbol{p}_{iA}(t,\boldsymbol{x})$ (A.4) of variable $A_i(t,\boldsymbol{x})$:

$$\boldsymbol{p}_{iA}(t,\boldsymbol{x}) = A_i(t,\boldsymbol{x})\boldsymbol{v}_i(t,\boldsymbol{x}) \tag{A.4}$$

We define collective flow $\boldsymbol{P}_A(t,\boldsymbol{x})$ of variable $A(t,\boldsymbol{x})$ as sum of agents flows (A.2, A.4) averaged during interval $\Delta$:

$$\boldsymbol{P}_A(t,\boldsymbol{x}) = \sum_{i\in dV(\boldsymbol{x});\,\Delta} A_i(t,\boldsymbol{x})\boldsymbol{v}_i(t,\boldsymbol{x}) = A(t,\boldsymbol{x})\boldsymbol{v}_A(t,\boldsymbol{x}) \tag{A.5}$$

Relations (A.5) allow present flow $\boldsymbol{P}_A(t,\boldsymbol{x})$ as product of variable $A(t,\boldsymbol{x})$ (A.1) and collective velocity $\boldsymbol{v}_A(t,\boldsymbol{x})$. Integral of (A.5) by (2.1, 2.2) define macroeconomic flow $\boldsymbol{P}_A(t)$ as product of variable $A(t)$ (A.3) and macroeconomic velocity $\boldsymbol{v}_A(t)$ in economic domain as:

$$\boldsymbol{P}_A(t) = \int d\boldsymbol{x}\; \boldsymbol{P}_A(t,\boldsymbol{x}) = A(t)\boldsymbol{v}_A(t) \tag{A.6}$$

Function $A(t,\boldsymbol{x})$ permit introduce mean risk $\boldsymbol{X}_A(t)$ as risk of all economic agents weighted by variable $A_i(t,\boldsymbol{x})$ and averaged during interval $\Delta$. Sum in (A.7) is taken by all agents $i \in E$ in economy E

$$\boldsymbol{X}_A(t)A(t) = \int d\boldsymbol{x}\; \boldsymbol{x}\, A(t,\boldsymbol{x}) = \sum_{i\in E;\,\Delta} \boldsymbol{x}\, A_i(t,\boldsymbol{x}) \tag{A.7}$$

**Appendix B**

## Market transactions, their flows and mean risks

We describe trades between agents. Agent $i$ at point $\boldsymbol{x}$ sells the volume $U_{ij}$ of variable $A$ to agent $j$ at point $\mathbf{y}$ and agent $j$ pays the value $C_{ij}$ to agent $i$. As a variable $A$ one may consider commodities, credits, investment, assets, service and etc. We define transaction $\boldsymbol{bs}_{ij}(t,\boldsymbol{x},\boldsymbol{y})$ between agents $i$ and $j$ at points $\boldsymbol{x}$ and $\boldsymbol{y}$, $z=(\boldsymbol{x},\boldsymbol{y})$ as:

$$\boldsymbol{bs}_{ij}(t,\boldsymbol{z}) = \left(U_{ij}(t,\boldsymbol{z}); C_{ij}(t,\boldsymbol{z})\right) \quad ; \quad \boldsymbol{z} = (\boldsymbol{x},\boldsymbol{y}) \tag{B.1}$$

We move from description of transactions as relations between agents and introduce transactions as functions of point $z=(\boldsymbol{x},\boldsymbol{y})$ in economic domain (3.2, 3.3). We define collective transaction $\boldsymbol{BS}(t,z)$ as sum of transactions between of agents $i$ inside small volume $dV(\boldsymbol{x})$ and agents $j$ inside volume $dV(\boldsymbol{y})$ (3.4) and average the sum during interval $\Delta$ alike to (A.1, A.2):

$$\boldsymbol{BS}(t,\boldsymbol{z}) = \left(U(t,\boldsymbol{z}); C(t,\boldsymbol{z})\right) \quad ; \quad \boldsymbol{z} = (\boldsymbol{x},\boldsymbol{y}) \tag{B.2}$$

$$U(t,\boldsymbol{z}) = \sum_{i\in dV(\boldsymbol{x}); j\in dV(\boldsymbol{y}); \Delta} U_{ij}(t,\boldsymbol{z}) \quad ; \quad C(t,\boldsymbol{z}) = \sum_{i\in dV(\boldsymbol{x}); j\in dV(\boldsymbol{y}); \Delta} C_{ij}(t,\boldsymbol{z}) \tag{B.3}$$

$$\sum_{i\in dV(\boldsymbol{x}); \Delta} U_{i,j}(t,\boldsymbol{z}) \equiv \frac{1}{\Delta}\int_{t-\Delta/2}^{t+\Delta/2} d\tau \sum_{i\in dV(\boldsymbol{x}); j\in dV(\boldsymbol{y})} U_{i,j}(\tau,\boldsymbol{z}) \tag{B.4}$$

Economic activity of sellers and buyers, evolution of economic environment change risk coordinates of agents involved into transactions. That causes flows of transactions alike to flows of economic variables (A.4-A.6). Seller $i$ with risk velocity $\boldsymbol{v}_{ix}$ carries flow $\boldsymbol{p}_{Uijx}$ of volume $U_{ij}(t,z)$ along axis $X$ and buyer $j$ with risk velocity $\boldsymbol{v}_{jy}$ carries flow $\boldsymbol{p}_{Uijx}$ volume $U_{ij}(t,z)$ along axis $Y$. Motion of agents $i$ and $j$ along axes $X$ and $Y$ define the flows of volume $U_{ij}(t,z)$ and value $C_{ij}(t,z)$ as

$$\boldsymbol{p}_{ij}(t,\boldsymbol{z}) = \left(\boldsymbol{p}_{Uij}(t,\boldsymbol{z}), \boldsymbol{p}_{Cij}(t,\boldsymbol{z})\right) \tag{B.5}$$

$$\boldsymbol{p}_{Uij}(t,\boldsymbol{z}) = \left(\boldsymbol{p}_{Uijx}(t,\boldsymbol{z});\ \boldsymbol{p}_{Uijy}(t,\boldsymbol{z})\right) \ ; \ \boldsymbol{p}_{Cij}(t,\boldsymbol{z}) = \left(\boldsymbol{p}_{Cijx}(t,\boldsymbol{z});\ \boldsymbol{p}_{Cijy}(t,\boldsymbol{z})\right) \tag{B.6}$$

$$\boldsymbol{p}_{Uijx}(t,\boldsymbol{z}) = U_{i,j}(t,\boldsymbol{z})\boldsymbol{v}_{ix}(t,\boldsymbol{x}) \quad ; \quad \boldsymbol{p}_{Uijy}(t,\boldsymbol{z}) = U_{i,j}(t,\boldsymbol{z})\boldsymbol{v}_{jy}(t,\boldsymbol{y}) \tag{B.7}$$

$$\boldsymbol{p}_{Cijx}(t,\boldsymbol{z}) = C_{i,j}(t,\boldsymbol{z})\boldsymbol{v}_{ix}(t,\boldsymbol{x}) \quad ; \quad \boldsymbol{p}_{Cijy}(t,\boldsymbol{z}) = C_{i,j}(t,\boldsymbol{z})\boldsymbol{v}_{jy}(t,\boldsymbol{y}) \tag{B.8}$$

To define collective flows $\boldsymbol{P}(t,z)$ and collective velocities $\boldsymbol{v}(t,z)$ of transactions as functions of coordinates averaged during time interval $\Delta$ we introduce the procedure similar to (B.2-B.4):

$$\boldsymbol{P}(t,\boldsymbol{z}) = \left(\boldsymbol{P}_U(t,\boldsymbol{z}), \boldsymbol{P}_C(t,\boldsymbol{z})\right) \ ; \ \boldsymbol{z} = (\boldsymbol{x},\boldsymbol{y}) \tag{B.9}$$

$$\boldsymbol{P}_U(t,\boldsymbol{z}) = U(t,\boldsymbol{z})\boldsymbol{v}_U(t,\boldsymbol{z}) \ ; \ \boldsymbol{P}_C(t,\boldsymbol{z}) = C(t,\boldsymbol{z})\boldsymbol{v}_C(t,\boldsymbol{z}) \tag{B.10}$$

$$\boldsymbol{P}_U(t,\boldsymbol{z}) = \left(\boldsymbol{P}_{xU}(t,\boldsymbol{z}); \boldsymbol{P}_{yU}(t,\boldsymbol{z})\right) \ ; \ \boldsymbol{P}_C(t,\boldsymbol{z}) = \left(\boldsymbol{P}_{xC}(t,\boldsymbol{z}); \boldsymbol{P}_{yC}(t,\boldsymbol{z})\right) \tag{B.11}$$

$$\boldsymbol{P}_{xU}(t,\boldsymbol{z}) = U(t,\boldsymbol{z})\boldsymbol{v}_{xU}(t,\boldsymbol{z}) = \sum_{i\in dV(\boldsymbol{x}); j\in dV(\boldsymbol{y})\,\Delta} U_{ij}(t,\boldsymbol{z})\boldsymbol{v}_{ix}(t,\boldsymbol{x}) \tag{B.12}$$

$$\boldsymbol{P}_{yU}(t,\boldsymbol{z}) = U(t,\boldsymbol{z})\boldsymbol{v}_{yU}(t,\boldsymbol{z}) = \sum_{i\in dV(\boldsymbol{x}); j\in dV(\boldsymbol{y})\,\Delta} U_{ij}(t,\boldsymbol{z})\boldsymbol{v}_{jy}(t,\boldsymbol{y}) \tag{B.13}$$

$$P_{xC}(t,z) = C(t,z)v_{xC}(t,z) = \sum_{i\in dV(x);j\in dV(y)\,\Delta} C_{ij}(t,z)v_{ix}(t,x) \quad (B.14)$$

$$P_{yC}(t,z) = C(t,z)v_{yC}(t,z) = \sum_{i\in dV(x);j\in dV(y)\,\Delta} C_{ij}(t,z)v_{jy}(t,y) \quad (B.15)$$

$$v(t,z) = \left(v_U(t,z);\ v_C(t,z)\right) \quad (B.16)$$

$$v_U(t,z) = \left(v_{xU}(t,z)\,;v_{yU}(t,z)\right)\ ;\ v_C(t,z) = \left(v_{xC}(t,z)\,;v_{yC}(t,z)\right) \quad (B.17)$$

Integrals by coordinates of transactions define total sales ***S(t,x)*** at point $\boldsymbol{x}$, total purchases at point $\boldsymbol{y}$ and total trades *BS(t)* in economy:

$$S(t,x) = \left(U_S(t,x); C_S(t,x)\right)\ ;\quad B(t,y) = \left(U_B(t,y); C_B(t,y)\right) \quad (B.18)$$

$$U_S(t,x) = \int dy\, U(t,x,y)\ ;\ C_S(t,x) = \int dy\, C(t,x,y) \quad (B.19)$$

$$U_B(t,y) = \int dx\, U(t,x,y)\ ;\ C_B(t,y) = \int dx\, C(t,x,y) \quad (B.20)$$

$$BS(t) = (Q(t); C(t)) \quad (B.21)$$

$$U(t) = \int dxdy\, U(t,x,y)\quad ;\quad C(t) = \int dxdy\, C(t,x,y) \quad (B.22)$$

Similar integrals of transactions define of total sellers flows $P_S(t,x)$ at point $\boldsymbol{x}$, total purchases flows $P_B(t,y)$ and total flows ***P(t)*** of transaction *BS(t)* in the economy. To avoid excess formulas we present definitions of total flows ***P(t)*** of transaction *BS(t)* in the economy

$$P(t) = (P_U(t); P_C(t)) \quad (B.23)$$

$$P_U(t) = U(t)v_U(t) = \int dz\ U(t,z)v_U(t,z) \quad (B.24)$$

$$P_C(t) = C(t)v_C(t) = \int dz\ C(t,z)v_C(t,z) \quad (B.25)$$

$$P_U(t) = \left(P_{xU}(t); P_{yU}(t)\right)\ ;\quad P_C(t) = \left(P_{xC}(t); P_{yC}(t)\right) \quad (B.26)$$

$$P_{xU}(t) = U(t)v_{xU}(t) = \int dz\ U(t,z)v_{xU}(t,z) \quad (B.27)$$

$$P_{yU}(t) = U(t)v_{yU}(t) = \int dz\ U(t,z)v_{yU}(t,z) \quad (B.28)$$

$$P_{xC}(t) = C(t)v_{xC}(t) = \int dz\ C(t,z)v_{xC}(t,z) \quad (B.29)$$

$$P_{yC}(t) = C(t)v_{yC}(t) = \int dz\ C(t,z)v_{yC}(t,z) \quad (B.30)$$

$$v(t) = (v_U(t)\,; v_C(t)\,) \quad (B.31)$$

$$v_U(t) = \left(v_{xU}(t)\,; v_{yU}(t)\right)\ ;\ v_C(t) = \left(v_{xC}(t)\,; v_{yC}(t)\right) \quad (B.32)$$

Transactions ***BS(t,z)*** (B.2-B.4) allow introduce mean risk $X_U(t)$ of volume *U(t)* and mean risk $X_C(t)$ of value *C(t)* similar to definitions of mean risk $X_A(t)$ of variable *A(t)* (A.7). We define mean risk $X_{BS}(t)$ of transaction ***BS(t)*** as two component function:

$$X_{BS}(t) = \left(X_U(t)X_U(t)\right) \quad (B.33)$$

$$X_U(t) = \left(X_{xU}(t)X_{yU}(t)\right)\ ;\quad X_C(t) = (X_{xC}(t)X_{yC}(t)) \quad (B.34)$$

$$X_{xU}(t)U(t) = \int dxdy\ x\, U(t,x,y)\ ;\quad X_{yU}(t)U(t) = \int dxdy\ y\, U(t,x,y) \quad (B.35)$$

$$X_{xC}(t)C(t) = \int dxdy\ x\, C(t,x,y)\ ;\quad X_{yC}(t)C(t) = \int dxdy\ y\, C(t,x,y) \quad (B.36)$$

Function $X_{xU}(t)$ – describes mean risk of sellers of volume $U(t)$ of trades $\boldsymbol{BS}(t)$ along axis **X** and $X_{yU}(t)$ – describes mean risk of buyers of the volume $U(t)$ along axis **Y** of economic domain (3.2, 3.3). $X_{xC}(t)$ and $X_{yC}(t)$ – describe mean risks of sellers and buyers of the value $C(t)$ along axes $\boldsymbol{X}$ and $\boldsymbol{Y}$. Economic domain (3.2, 3.3) is a unit cube with $2n$-dimension and hence any motions of mean risks (B.33-B.36) can be presented as fluctuations of mean risks from secure area of economic domain to risky area and back. These fluctuations of mean risks illustrate the phases of business cycles.

**Appendix C**

## The model of supply-demand cycles

We study the cycles of macroeconomic supply $S(t)$ and demand $D(t)$ under action of single risk in the 1-dimensional economic domain [0,1]. We present model of self-consistent interaction between small dimensionless disturbances of supply $s(t,x)$ and demand $d(t,x)$ near stationary state in the linear approximations by perturbations. We take supply $S(t,x)$ and demand $D(t,x)$:

$$S(t,x) = S_0\big(1 + s(t,x)\big)\ ; \quad D(t,x) = D_0\big(1 + d(t,x)\big) \tag{C.1}$$

$$|s(t,x)| \ll 1 \ \ ; \ \ |d(t,x)| \ll 1 \quad ; \quad 0 \le x \le 1$$

Supply $S(t,x)$ and demand $D(t,x)$ and their flows $P_S(t,x)=S(t,x)\upsilon_s(t,x)$ and $P_D(t,x)=D(t,x)\upsilon_d(t,x)$ follow equations similar to (4.5, 4.6):

$$\frac{\partial}{\partial t}S(t,x) + \nabla\cdot\big(S(t,x)v_s(t,x)\big) = F_D(t,x)\ ;\ \frac{\partial}{\partial t}D(t,x) + \nabla\cdot\big(D(t,x)v_d(t,x)\big) = F_S(t,x) \tag{C.2}$$

$\frac{\partial}{\partial t}P_S(t,x) + \nabla\cdot\big(P_S(t,x)v_s(t,x)\big) = F_{PD}(t,x)\,;\ \frac{\partial}{\partial t}P_D(t,x) + \nabla\cdot\big(P_D(t,x)v_d(t,x)\big) = F_{PS}(t,x)$ (C.3)

We propose that velocities of supply $\upsilon_s(t,x)$ and demand $\upsilon_d(t,x)$ are small. We assume that interactions between supply and demand in linear approximation by disturbances take form:

$$F_D(t,x) \sim aD_0\, xv_d(t,x)\ \ ;\ \ F_S(t,x) \sim bS_0\, x\, v_s(t,x) \tag{C.4}$$

$$F_{PD}(t,x) \sim \alpha D_0 v_d(t,x)\ \ ;\ \ F_{PS}(t,x) \sim \beta S_0 v_s(t,x) \tag{C.5}$$

We neglect the square terms by $\upsilon_s^2$ and $\upsilon_s^2$ in (C.3) and in the linear approximation by disturbances equations (C.1-C.5) on $s(t,x)$ and $d(t,x)$ and velocities $\upsilon_s(t,x)$, $\upsilon_d(t,x)$ take form:

$$S_0\frac{\partial}{\partial t}s(t,x) + S_0\frac{\partial}{\partial x}v_s(t,x) = aD_0\, xv_d(t,x) \tag{C.6}$$

$$D_0\frac{\partial}{\partial t}d(t,x) + D_0\frac{\partial}{\partial x}v_d(t,x) = bS_0\, x\, v_s(t,x) \tag{C.7}$$

$$S_0\frac{\partial}{\partial t}v_s(t,x) = \alpha D_0 v_d(t,x) \quad ; \quad D_0\frac{\partial}{\partial t}v_d(t,x) = \beta S_0 v_s(t,x) \tag{C.8}$$

Equation (C.6) describes dimensionless supply disturbances *s(t,x)* with right side determined by 1-dimensional scalar product of risk coordinate *x* and demand flow $D_0 v_d(t,x)$. This models the impact of the demand flow in economic domain on evolution of supply disturbances *s(t,x)*. If $v_d(t,x)>0$ then right side (C.6) describes demand flow in the direction of risk growth that cause growth of supply *s(t,x)*. Negative $v_d(t,x)<0$ models demand risk aversion and cause decline of supply *s(t,x)*. Equation (C.7) models similar impact of the 1-dimensional scalar product of risk coordinate *x* and supply flow $S_0 v_s(t,x)$ on demand disturbances *d(t,x)*. Right side in (C.6; C.7) models direct impact of risk coordinate *x* on supply and demand. Equation (C.7) describes linear mutual relations between supply and demand flows. Positive demand flow with $v_d(t,x)>0$ cause growth of supply flow $S_0 v_s(t,x)$ and negative demand flow with $v_d(t,x)<0$ results in decline of supply velocity $v_s(t,x)$. Coefficient $\alpha>0$ describes the effect of these relations. Equation of demand flow (C.7) models the similar relations but with the opposite sign of the coefficient $\beta<0$. Growth of supply flow $S_0 v_s(t,x)$ reduce demand *d(t,x)* and decline of supply flow cause increase of demand. Equations (C.8) describe mutual dependence between velocities of supply and demand in the linear approximation by disturbances. We show below that equations (C.1; C.6-C.8) describe the cycles of supply *s(t)* and demand *d(t)* that are accompanied by fluctuations of mean risk of supply $X_s(t)$ and mean risk of demand $X_d(t)$.

It is easy to show that integrals *dx* of (C.6-C.8) by economic domain [0,1] give:

$$S_0 \frac{d}{dt} s(t) = aD_0 \int dx \; x v_d(t,x) = \; aD_0 v_{dx}(t) \tag{C.9}$$

$$D_0 \frac{d}{dt} d(t) = bS_0 \int dx \; x v_s(t,x) = \; bS_0 v_{sx}(t) \tag{C.10}$$

$$S_0 \frac{d}{dt} v_s(t) = \alpha D_0 v_d(t) \quad ; \quad D_0 \frac{d}{dt} v_d(t) = \beta S_0 v_s(t) \tag{C.11}$$

$$s(t) = \int dx \; s(t,x) \; ; \; \; d(t) = \int dx \; d(t,x) \; ;$$

$$v_s(t) = \int dx \; v_s(t,x) \; ; \; \; v_d(t) = \int dx \; v_d(t,x)$$

For

$$\omega^2 = -\alpha\beta > 0 \tag{C.12}$$

equations (C.11) describe harmonic oscillations of velocities $v_d(t)$, $v_d(t)$ with frequency $\omega$ (C.12):

$$v_s(t) = v_{s0} \sin \omega t \quad ; \quad v_{dx}(t) = \frac{\omega S_0}{\alpha D_0} \, v_{s0} \cos \omega t \tag{C.13}$$

Let's multiply both parts in (C.8) by x and take integral over domain [0,1]. That gives equations on functions $v_{sx}(t)$ and $v_{dx}(t)$ that describe fluctuations with frequency (C.12):

$$v_{sx}(t) = v_{sx0}\sin\omega t \quad ; \quad v_{dx}(t) = \frac{\omega S_0}{\alpha D_0}\, v_{sx0}\cos\omega t \tag{C.14}$$

Functions (C.14) define simple solutions for (C.9; C.10) as:

$$s(t) = \frac{a}{\alpha} v_{sx0}\sin\omega t \quad ; \quad d(t) = -b\frac{S_0}{\omega D_0} v_{sx0}\cos\omega t \tag{C.15}$$

(C.15) describes cycles of supply *s(t)* and demand *d(t)* disturbances with the same frequency (C.12) as oscillations of the supply and demand flows $S_0 v_s(t)$ and $D_0 v_d(t)$ (C.13). To derive fluctuations of mean risk let's define mean supply risk $X_s(t)$ and mean demand risk $X_d(t)$ as:

$$S(t)X_s(t) = \int_0^1 dx\, x\, S(t,x) = S_0\int_0^1 dx\;\, x\big(1 + s(t,x)\big) = S_0\left(\frac{1}{2} + f(t)\right) \tag{C.16}$$

$$D(t)X_d(t) = \int_0^1 dx\, x\, D(t,x) = D_0\int_0^1 dx\;\, x\big(1 + d(t,x)\big) = D_0\left(\frac{1}{2} + g(t)\right) \tag{C.17}$$

$$f(t) = \int_0^1 dx\;\, xs(t,x) \quad ; \quad g(t) = \int_0^1 dx\;\, xd(t,x)$$

Let's present mean supply risk $X_s(t)$ and mean demand risk $X_d(t)$ as:

$$X_s(t) = \frac{1}{2}\big(1 + x_s(t)\big) \quad ; \quad X_d(t) = \frac{1}{2}\big(1 + x_d(t)\big) \tag{C.18}$$

Disturbances $x_s(t)$ and $x_d(t)$ describe small fluctuations of mean risks $X_s(t)$ and $X_d(t)$ near ½.In linear approximation by disturbances *s(t), d(t),* $x_s(t)$ and $x_d(t)$ (C.1, C.16, C.17, C.18) give:

$$x_s(t) = 2f(t) - s(t) \quad ; \quad x_d(t) = 2g(t) - d(t) \tag{C.19}$$

To derive equations of functions *f(t)* and *g(t)* let's multiply (C.6, C.7) by x and take integral.

$$S_0\frac{d}{dt}f(t) = S_0 v_s(t) + aD_0 v_{dx2}(t) \quad ; \quad v_{dx2}(t) = \int_0^1 dx\;\, x^2 v_d(t,x) \tag{C.20}$$

$$D_0\frac{d}{dt}g(t) = D_0 v_d(t) + bS_0 v_{sx2}(t) \quad ; \quad v_{sx2}(t) = \int_0^1 dx\;\, x^2 v_s(t,x) \tag{C.21}$$

To derive equations on functions $v_{sx2}(t)$ and $v_{dx2}(t)$ let's multiply (C.8) by $x^2$ and take integrals. It is obvious that functions $v_{sx2}(t)$ and $v_{dx2}(t)$ follow similar harmonic oscillations with frequency (C.12).

$$v_{sx2}(t) = v_{sx20}\sin\omega t \quad ; \quad v_{dx}(t) = \frac{\omega S_0}{\alpha D_0}\, v_{sx20}\cos\omega t \tag{C.22}$$

Relations (C.13, C.15, C.22) give solutions for (C.20, C.21) and present risk disturbances $x_s(t)$ and $x_d(t)$ (C.19) as:

$$x_s(t) = \frac{a}{\alpha}\, v_{sx20}\sin\omega t - 2\frac{v_{s0}}{\omega}\cos\omega t \tag{C.23}$$

$$x_d(t) = 2\frac{S_0}{\alpha D_0}\, v_{s0}\sin\omega t - b\frac{S_0}{\omega D_0}(2v_{sx20} - v_{sx0})\cos\omega t \tag{C.24}$$

Thus small disturbances of supply *s(t)* and demand *d(t)* (C.15) and disturbances of mean risks of supply $x_s(t)$ and demand $x_d(t)$ fluctuate with the same frequency (C.12) determined by fluctuations of supply $S_0 v_s(t)$ and demand $D_0 v_d(t)$ flows (C.11)